%% file: main.tex
\documentclass[manuscript]{acmart}
\AtBeginDocument{%
  }

\setcopyright{acmlicensed}
\copyrightyear{2026}
\acmYear{2026}
\acmDOI{XXXXXXX.XXXXXXX}
\acmConference[ICER '26]{International Computing Education Research}{August 11--14,
  2026}{Uppsala, Sweden}
\acmISBN{978-1-4503-XXXX-X/2018/06}

\usepackage[table]{xcolor} 
\usepackage{multirow}

\usepackage{enumerate}
\usepackage{enumitem}
\usepackage{fdsymbol}
\usepackage{subcaption}
\usepackage{todonotes}
\usepackage{pgfplotstable,booktabs}
\usetikzlibrary{positioning}
\usepackage[dvipsnames]{xcolor}
\usepackage{pgfplots}
\pgfplotsset{compat=1.18} 
\usepackage{multirow}
\usepackage{makecell}
\usepackage{tabularx}
\usepackage[table]{xcolor}
\usepackage{hyperref}
\usepackage{xparse}
\usepackage{array}

\newcolumntype{s}{>{\centering\arraybackslash}X}
\newcolumntype{B}{>{\raggedright\arraybackslash}X}

\pgfplotstableread{
Venue Total Replication
Conferences 1939 47
Journals 330 7
}\datatableunsort

\newlength\WIDTHOFBAR
\definecolor{nonreplication}{RGB}{0,69,122}
\definecolor{replication}{RGB}{253,177,26}
\def\shownonreplicationbar#1{%
  \def\temp{#1}%
  \ifx\temp\empty
    \def\barval{0}%
  \else
    \def\barval{#1}%
  \fi
  {\color{nonreplication!100}\rule{\barval cm}{6pt}}%
  {\color{nonreplication!100}\rule{\WIDTHOFBAR - \barval cm}{6pt}}%
}
\def\showreplicationbar#1{%
  \def\temp{#1}%
  \ifx\temp\empty
    \def\barval{0}%
  \else
    \def\barval{#1}%
  \fi
  {\color{replication!100}\rule{\barval cm}{6pt}}%
  {\color{replication!100}\rule{\WIDTHOFBAR - \barval cm}{6pt}}%
}

\def\undergradbar#1{%
  \def\temp{#1}%
  \ifx\temp\empty
    \def\barval{0}%
  \else
    \def\barval{#1}%
  \fi
  {\color{nonreplication!100}\rule{\barval cm}{6pt}}%
  {\color{nonreplication!100}\rule{\WIDTHOFBAR - \barval cm}{6pt}}%
}

\def\csonebar#1{%
  \def\temp{#1}%
  \ifx\temp\empty
    \def\barval{0}%
  \else
    \def\barval{#1}%
  \fi
  {\color{replication!100}\rule{\barval cm}{6pt}}%
  {\color{replication!100}\rule{\WIDTHOFBAR - \barval cm}{6pt}}%
}

\def\elembar#1{%
  \def\temp{#1}%
  \ifx\temp\empty
    \def\barval{0}%
  \else
    \def\barval{#1}%
  \fi
  {\color{SpringGreen!100}\rule{\barval cm}{6pt}}%
  {\color{SpringGreen!100}\rule{\WIDTHOFBAR - \barval cm}{6pt}}%
}

\def\gradbar#1{%
  \def\temp{#1}%
  \ifx\temp\empty
    \def\barval{0}%
  \else
    \def\barval{#1}%
  \fi
  {\color{WildStrawberry!100}\rule{\barval cm}{6pt}}%
  {\color{WildStrawberry!100}\rule{\WIDTHOFBAR - \barval cm}{6pt}}%
}

\def\otherbar#1{%
  \def\temp{#1}%
  \ifx\temp\empty
    \def\barval{0}%
  \else
    \def\barval{#1}%
  \fi
  {\color{Goldenrod!100}\rule{\barval cm}{6pt}}%
  {\color{Goldenrod!100}\rule{\WIDTHOFBAR - \barval cm}{6pt}}%
}

\makeatletter
\newcommand\footnoteref[1]{\protected@xdef\@thefnmark{\ref{#1}}\@footnotemark}
\makeatother

\definecolor{lightgray}{gray}{0.9} 

\begin{document}

\title{Revisiting the Replication Study Design Used in Computing Education Research}


\author{Rita Garcia}
\orcid{0000-0003-4615-4921}
\email{rita.garcia@vuw.ac.nz}
\affiliation{%
  \institution{Victoria University of Wellington}
  \state{}
  \country{Wellington, New Zealand}
}

\author{Ellie Lovellette}
\orcid{0000-0002-6944-5206}
\affiliation{%
  \institution{College of Charleston}
  \state{South Carolina}
  \country{USA}
}
\email{lovelletteeb@cofc.edu}

\author{Xi Wu}
\orcid{0000-0001-5795-9798}
\affiliation{%
  \institution{University of Sydney}
  \country{Australia}
}
\email{xi.wu@sydney.edu.au}

\author{Angela Zavaleta Bernuy}
\orcid{0000-0002-1228-5774}
\email{zavaleta@mcmaster.ca}
\affiliation{
  \institution{McMaster University}
  \country{Hamilton, ON, Canada}
}

\renewcommand{\shortauthors}{Garcia et al.}

\begin{abstract}
\input{00abstract}
\end{abstract}



\maketitle

\input{01introduction}
\input{02background}
\input{03relatedwork}
\input{04studydesign}
\input{05results}
\input{06discussion}
\input{07limitations}
\input{08conclusion}
\bibliographystyle{ACM-Reference-Format}
\bibliography{sample-base}


\end{document}

%% file: 00abstract.tex
\textbf{Background and Context:} Replication studies play an important role in Computing Education Research (CER) by supporting the development of consistent and reliable scientific knowledge. However, prior research indicates that the CER community tends to prioritise novel contributions over replication. A 2019 Systematic Literature Review (SLR) identified only 54 (2.38\%) replication studies among 2,269 papers published between 2009 and 2018 across five major CER venues. In response, the Computer Science Education journal released a special issue dedicated to replication studies to encourage greater adoption of this research design.
\textbf{Objectives:} This study aims to examine how the landscape of replication research in CER has evolved since 2019. Specifically, we investigate whether the prevalence of replication studies has increased and explore current perceptions and experiences of CER researchers regarding replication.
\textbf{Method:} We replicated two prior studies. First, we conducted an updated SLR to identify replication studies published between 2019 and 2025 in the same five CER venues. Second, we replicated a survey of Computing Education researchers to better understand their perceptions, experiences, and challenges related to conducting and publishing replication studies.
\textbf{Findings:} Our SLR identified 63 (2.50\%) replication studies among 2,516 published papers. While the proportion of replication studies has increased slightly, overall growth remains limited.  We observed a shift toward more published replication studies in journals and an increase in authors replicating their own prior work. Survey results indicate that although many researchers engage in replication within their teaching and research practice, they encounter significant challenges when attempting to publish replication studies.
\textbf{Implications:} Despite increased discourse around open science and research rigour, the adoption of replication studies in CER has not substantially grown. Our findings offer opportunities for future research to promote replication in CER and to explore how the CER community can encourage researchers to publish replication studies.

%% file: 01introduction.tex
\section{Introduction}\label{section_introduction}

A replication study is a research type with similar goals and study methods as another study, designed to confirm consistent conclusions and verify the reliability and validity of the original study \cite{jeffreys1974ReplicationDefinition}. The Computing Education Research (CER) community encourages researchers to do more replication studies because they ensure the original study is not a phenomenon and confirm prior findings in different contexts \cite{ahadi2016CERSurveyOnReplication, guzdial2016Miranda}. Unfortunately, replication studies are not commonly applied in CER \cite{hao2019ReplicationSLR} due to educational institutions \cite{romero2019replicability} and the community \cite{ahadi2016CERSurveyOnReplication, guzdial2016Miranda} placing a higher value on novel research. A 2019 Systematic Literature Review (SLR) \cite{hao2019ReplicationSLR} investigated the application of replication in CER, identifying 54 papers (2.38\%) published in five computing venues between 2009 and 2018. However, since the publication of the 2019 SLR, the Computer Science Education Journal (CSEJ) published a 2022 special issue\footnote{\label{CSEJ}https://www.tandfonline.com/toc/ncse20/32/3} focusing on replication studies, which motivated us to re-examine the adoption of replication studies in CER over the last seven years. For this study, we address the following research questions (RQs):

\begin{enumerate}
    \item \textbf{RQ1:} How has the adoption of replication study design changed over the last seven years?
    \item \textbf{RQ2:} What perceptions do Computing Education researchers have of using replication in their work?
    \item \textbf{RQ3:} How can the CER community support researchers in adopting the replication study design?
\end{enumerate}

To answer these RQs, we bring together two studies focusing on replication studies in the context of today's research landscape, where AI is available to assist students in solving activities and educators teaching post-COVID are facing challenges, such as encouraging student participation and engagement in distance learning environments \cite{neuwirth2021covidHigherEdu}. These factors merit revisiting research within these contexts. We replicated a 2019 SLR \cite{hao2019ReplicationSLR} to identify replication studies published between 2019 and 2025. Like the original study, we used the same approach to identify replication studies in five prominent Computing Education venues, enabling us to compare and contrast results with the original study. We also replicated a study \cite{ahadi2016CERSurveyOnReplication} that surveyed Computing Education researchers to collect their perspectives on the study design. We used a mixed-methods approach to analyse the collected data. 

Our SLR identifies 63 (2.50\%) papers among the 2516 published in the five venues during the time period of interest, with a shift toward publishing in journals and authors replicating their own work. We found replication has not grown as the current discourse around open science and rigour suggests \cite{Korbmacher2023}. However, when speaking with Computing Education researchers, many are actively engaging in replication in their own teaching and research, but are experiencing various challenges with this research design. From these findings, we identify opportunities to advance replication in CER and explore how the CER community can encourage researchers to publish replication studies successfully.


%% file: 02background.tex
\section{Background}\label{section_background}

Replication is a critical mechanism for enhancing scientific knowledge by assessing the reliability and validity of prior findings. At its core a \textit{replication study} uses similar goals and design methods as another study, to evaluate whether earlier conclusions remain valid \cite{jeffreys1974ReplicationDefinition}. Replication studies could also help determine whether results hold under new conditions, different populations, or alternative methodological choices. Prior research \cite{brown2022launching, dennis2014ReplicationManifesto, huffmeier2016ReplicationTypology, lykken1968StatSignPsyResearch, Ampel2023} presented the following replication types:
 
\begin{itemize}[leftmargin=*]
    \item \textbf{Exact replication} studies attempt to reproduce the original study's design and procedures under the same conditions.
    \item \textbf{Conceptual replication} studies investigate the same underlying hypothesis using different operational methods, measurements, or instructional contexts. 
    \item \textbf{Methodological replication} studies vary measurement tools or techniques to test the robustness of findings \cite{dennis2014ReplicationManifesto,Ampel2023}.
    \item \textbf{Operational (direct) replication} studies preserve the conditions with identical sampling and empirical procedures to reproduce findings outside of the original context.
\end{itemize}
Despite their importance, replication studies remain uncommon in the CER community. One barrier is that many publications do not provide sufficient methodological details for researchers to repeat the work reliably \cite{ihantola2015DataMiningLitReview}. A potential reason may be due to space limitations imposed by the publication venue.

Previously, \citet{brown2022launching} addressed publication bias in CER for replication research by launching registered reports that were peer-reviewed before data collection, ensuring methodological rigour. The initiative demonstrated the feasibility of registered report replications and highlighted how editorial intervention can aid diverse research methodologies. This effort led to the aforementioned CSEJ special issue\footnoteref{CSEJ} on replication studies, which only accepted registered reports replicating prior studies.  

In summary, replication is fundamental to enhancing the credibility and generalisability of scientific findings, yet its uptake within the CER community remains limited, partly due to inadequate methodological transparency and constraints imposed by publication venues. Recent initiatives have demonstrated that systematic replication, such as registered reports, can help address publication bias and promote methodological rigour. These efforts highlight the continuing need to refine editorial processes and prioritise replicability, ultimately strengthening the robustness and reliability of research in CER.

%% file: 03relatedwork.tex
\section{Related Work}\label{section_related_work}
Our work replicates two studies \cite{ahadi2016CERSurveyOnReplication, hao2019ReplicationSLR} focusing on replication studies in Computing Education Research (CER). We mentioned in \autoref{section_introduction} the 2019 \citet{hao2019ReplicationSLR} Systematic Literature Review (SLR) on replications, where the researchers used the query ``\textit{replicat[a-z]*}'' to identify papers. Of the 2,269 papers published between 2009 and 2018 across the five venues, that SRL identified 54 (2.38\%) replication studies and examined:

\begin{itemize}[leftmargin=*]
    \item Whether the papers performed a direct, conceptual, or other types of replication study design,
    \item Whether the findings are consistent with the replication target,
    \item Whether the same authors conducted the replication studies, ensuring one overlapping author,
    \item What methodologies the authors adopted, such as a quantitative, qualitative, or mixed-methods approach, and
    \item The applied themes, based on the 18 themes proposed in prior computing education research \cite{pears2005coreCERLiterature, sheard2009teachLearnProg, valentine2004CSERMetanalysis}.
\end{itemize}

When presenting the results, the researchers separated the journal and conference papers, stating, ``the two journals (TOCE and CSEJ) published 330 studies between 2009 and 2018, of which 7 studies were identified as replication studies. In comparison, the three conferences (SIGCSE, ICER, and ITiCSE) published 1,939 studies between 2009 and 2018, of which 47 studies were identified as replication studies'' \cite[p.~5]{hao2019ReplicationSLR}. The results showed that the majority of papers used the conceptual replication type, which tests the original hypothesis and theoretical values, but with different measures for data collection and analysis \cite{huffmeier2016ReplicationTypology}. The majority (N=34, 63\%) of the papers confirmed the findings of their original studies, predominantly in quantitative replication studies (N=40, 74\%). 

Overall, the replication studies identified by \citet{hao2019ReplicationSLR} predominantly focused on learning and teaching strategies, assessment, learning behaviour and theory, and performance prediction, with the majority (N=29, 54\%) focusing on undergraduate CS students. The authors argue for the need for replication in CER and acknowledge challenges and stigmas we also encountered in our conversations with CER researchers. The study suggested solutions to encourage more replication in CER, including revisions to policies to support replication and guidelines for reviewing replication studies. The authors concluded by promoting replication on merit, noting that ``a healthy environment of replication studies needs to be cultivated and maintained'' \cite[p.~11]{hao2019ReplicationSLR}.

In the 2016 work by \citet{ahadi2016CERSurveyOnReplication}, the researchers examined the CER community's views on replication studies. The authors surveyed 73 Computing Education researchers to collect their perceptions. Their findings showed that 
participants perceived that novel work had a greater impact and offered more opportunities for researchers to secure grants and citations. Replication work was acknowledged to be difficult, with a large portion of researchers attempting and failing to replicate their own results (48\%) or the result of another researcher (40\%). When participants were asked if they had published a replication study, 18\% of respondents answered in the affirmative, while 19\% had tried but were unsuccessful. The maximum number of published replications was 2, by two respondents.

These two studies provide insights into the landscape of replication in CER, highlighting both its limited presence and the challenges it poses. The SLR study \cite{hao2019ReplicationSLR} revealed that replication studies remain a small proportion of published work, with a majority confirming original findings. The survey study \cite{ahadi2016CERSurveyOnReplication} further underscored the difficulties researchers face in replicating studies, alongside participants' perceptions that the CER community valued novel research more in terms of impact and career progression. Both studies advocate cultivating a supportive environment for replication, suggesting policy changes and clearer guidelines to encourage and recognise its importance in CER, a goal we also seek to advance.

%% file: 04studydesign.tex
\section{Study Design}\label{section_study_design}

\begin{figure*}
\centering
  \includegraphics[width=0.65\textwidth]{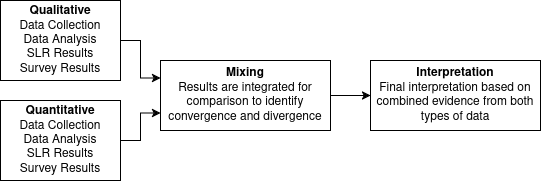}

    \caption{Diagram of Study Method using Triangulation Mixed-Methods Design (Adapted from \citet{creswell:2006})}
  \label{fig:study_method_figure}
\end{figure*}

Our study design first conducted a methodological replication study \cite{dennis2014ReplicationManifesto} of the 2019 Systematic Literature Review \cite{hao2019ReplicationSLR} we examined in \autoref{section_related_work}. We then performed a conceptual replication study \cite{huffmeier2016ReplicationTypology} of the \citet{ahadi2016CERSurveyOnReplication} survey study described in  the same section. We used a mixed-methods approach \citep{creswell:2012}, illustrated in \autoref{fig:study_method_figure}, employing a triangulation design \citep{creswell:2006} to analyse the collected data. We collected and analysed the data from the SLR and the survey separately, then integrated them to compare and identify areas of convergence, complementarity, or divergence. Figure \ref{fig:study_method_figure} shows how we bring together the findings from the SLR and survey, with the survey results offering additional perspectives on the SLR findings. We made our instruments, data collected from the SLR, and SLR references available on FigShare.\footnote{https://doi.org/10.6084/m9.figshare.28862405} We obtained ethics (IRB) approval from the ethics committee at {\color{purple}<<university>>} to conduct this study. 


\subsection{Participants}\label{subsection_participants}

We recruited survey participants through multiple channels. We distributed the survey to the SIGCSE members' mailing list and posted announcements in community forums and social networks. 
In addition, we used snowball sampling \cite{abdulquader2006SnowballResponse}, asking colleagues and contacts to take the survey and/or to distribute it to their own networks to increase response rates. 


\subsection{Systematic Literature Review}\label{subsubsection_systematic_literature_review}


For our replication of the 2019 SLR study \cite{hao2019ReplicationSLR}, we utilised the same methods as the original SLR, including the same venues, but for the subsequent years: 2019-2025. We applied the search query ``\textit{replicat[a-z]*}'' to the digital libraries for the five venues: SIGCSE, ICER, ITiCSE, Transactions of Computing Education (TOCE), and Computer Science Education Journal (CSEJ). We searched for the string in the entire paper, including the title, keywords, and abstract. Our search query identified 645 papers across the five venues with no duplicates.

Two raters conducted the first pass of the selection process to identify papers replicating other studies. To understand the original study's selection process, the raters reviewed papers \cite{frank2012teachReplication, makel2012replicationPsychology, makel2014replicationEduSciences} \citet{hao2019ReplicationSLR} used in the original study to construct the selection criteria. We examined 15 of their selected papers to better understand how the original authors met the selection criteria. In addition, we spoke with a co-author of the original study for further insight.

The raters used a spreadsheet for the selection process. If there was a discrepancy between their selections, a third rater examined and discussed the selections to reach a decision. The third rater intervened on four (0.16\%) papers during the selection process. For two papers, all three raters were unsure how to rate, so we contacted the papers' primary authors for confirmation. During the selection process, one rater recorded papers that encouraged replication of their work by explicitly stating it, making their research materials available for future research, or both. For example, one paper \cite{bart2019encourageReplicationExample} stated, ``to encourage replication and extension of our analysis, we make these scripts publicly available through a GitHub repository'' \cite[p.~176]{bart2019encourageReplicationExample}. Of the 645 papers collected from our search query, we identified 63 (2.50\%) actual replication studies according to the definition. We used Cohen's Kappa (k) to measure interrater reliability \cite{cohen1968kappa} in the selection process, showing that the raters' coding achieved a kappa of 0.952, which is considered almost perfect agreement \cite{landis1977agreement}.  

We analysed the 63 papers that met our selection criteria, using Google Forms to collect the data and store it in a spreadsheet, mirroring the approach used in the original study. 
We recorded the replication study types, the methodology, authors, and themes used in the replication papers. For the replication types, the form listed \textit{Direct} and \textit{Conceptual} based on the work by \citet{hao2019ReplicationSLR}, but included an open-text field for other replication types. 

We used comparative analysis \cite{pickvance:2001} to assess our results against the original SLR, focusing on context, methodologies, topics, and authorship. We used the quantified evidence presented in the original SLR to compare with our findings.

To help situate our SLR results, we identified the number of accepted papers in the venues. For the conferences, we used the conference chairs' welcome, which presents the number of accepted papers.\footnote{https://dl.acm.org/doi/proceedings/10.1145/3702652} For the journals, we used their online digital libraries to identify the total number of published papers for each year. Upon completing the data analysis, we extracted the frequency matrix from the spreadsheet to present our findings.



\subsection{Survey}\label{subsubsection_survey}

We conducted a conceptual replication study of the research by \citet{ahadi2016CERSurveyOnReplication}. The survey informs the SLR's findings and provides a more in-depth look at the CSER community's application of replication study design in their work. To adhere to appropriate ethics (IRB) requirements, we hosted the survey on the {\color{purple}<<university>>}'s Qualtrics platform, where the research group gained Institutional Review Board approval. 

The survey contained 12 questions: four closed-ended and eight open-ended. Depending on how the participant answered the first survey question asking them if they had done a replication study before, they were presented with different sets of questions. 

Participants who had previously performed replication studies received six additional questions focusing on their prior experience with those studies, for example, \textit{``How many CER replication studies have you conducted?''} and \textit{``How many CER replication studies have you successfully published?''}. We also asked those participants to provide any replication studies they had previously published, provided they did not mind potential deanonymisation by the research team. If the respondent provided their published replication studies, two of the authors downloaded and reviewed them to determine the publication year, venue, replication type, and whether the publication appeared in the original or our SLR. Analysing the provided papers allowed us to determine whether the CER community published replication studies outside the five venues and to examine whether our selection criteria would have selected them for review, thereby informing how we examine and identify replication studies in CER in the future. 

All participants, regardless of the answer they gave to the first question, were asked to provide their opinions on replication studies with the open-text question \textit{``What are your opinions on the value of replication studies in CER?''}. We asked if they were planning on conducting a replication study in the future, and gave them the chance to write am open-ended response to the question \textit{``Why do you think the CER community does not do more replication studies?''}. 

Two authors independently reviewed and coded the open-ended responses, each applying their own preliminary coding scheme. After completing this initial coding round, they met to compare their code assignments and collaboratively develop a unified codebook, drawing on their initial tags and interpretations. Using the agreed-upon codebook, both authors re-coded the responses to ensure consistency with the standardised codes. One week later, they met again to review their coding, discuss any discrepancies, and reconcile differences. They repeated this process until they reached consensus on all codes, thereby enhancing the reliability and validity of the coding procedure for the dataset.





\subsection{Positionality Statement}

Our research team consists of four women researchers, one identifying as white and three identifying as members of different minoritised racial groups. We adopt a mixed-methods perspective and acknowledge that our backgrounds and experiences shape our interpretations of replication practices in CER \cite{braun2023toward}. To mitigate individual bias, we employed collaborative and reflexive analytic practices throughout the study. Two researchers independently conducted paper selection for the Systematic Literature Review, with disagreements resolved through discussion and third-rater adjudication, informed by the procedures of the original study. For the survey’s open-ended responses, two authors independently coded the data, developed a shared codebook, and iteratively reconciled differences until they reached consensus. These practices were intended to strengthen analytic rigour and balance individual perspectives when interpreting replication types and community attitudes toward replication in CER.

%% file: 05results.tex
\section{Results}\label{section_results}


\subsection{Systematic Literature Review Results}\label{subsection_slr_results}

Of the 2516 papers published in SIGCSE, ITiCSE, ICER, TOCE and CSEJ between 2019 and 2025, 63 (2.50\%) applied replication. Of these 2516 papers, 147 (6\%) encouraged replication of their work. \autoref{tab:slr_results} presents our SLR results, showing the total number of publications per venue and the replication studies identified at each venue. The table groups the results by conference and journal, with years presented in descending order. It also shows the total number of papers and replication studies for each venue within the selected time frame. 

\input{tables/slr_yearly_results}

When examining results by venue and year, CSEJ 2022 has the highest percentage of replication studies (N=4, 21\%), mainly because of the CSEJ special issue focusing on replication studies published that year, denoted with \dagger \space in \autoref{tab:slr_results}. TOCE 2025 has the next-highest number of replication studies (N=6, 7\%). We also found nine journal volumes or conference proceedings containing no replication studies at all, denoted with \ddagger \space in \autoref{tab:slr_results}. These include CSEJ 2020, 2021, 2023, and 2025; TOCE 2020 and 2021; ICER 2022; and ITiCSE 2019 and 2025. When examining the number of replication studies published as a percentage of all publications for the time period at each venue, we found that ICER published the most (5\%, N=11), followed by CSEJ (4\%, N=6), while SIGCSE (2\%, N=20) had the fewest.

\begin{figure}[b]
    \begin{subfigure}[b]{\textwidth}
        \centering
        \resizebox{.95\linewidth}{!}{\input{figures/conference_stackedbar.tex}}
        \caption{Conference Results}\label{fig:conference_stackbar}
    \end{subfigure}
    \begin{subfigure}[b]{\textwidth}
        \centering
        \resizebox{.95\linewidth}{!}{\input{figures/journal_stackedbar.tex}}
        \caption{Journal Results}\label{fig:journal_stackbar}
    \end{subfigure}
\caption{Comparing Original and Replication Studies' Results by Conference and Journal Publications}\label{fig:slr_paper_comparison}
\vspace{2mm}
\shownonreplicationbar{} Non-Replication Papers \showreplicationbar{} Replication Papers
\end{figure}
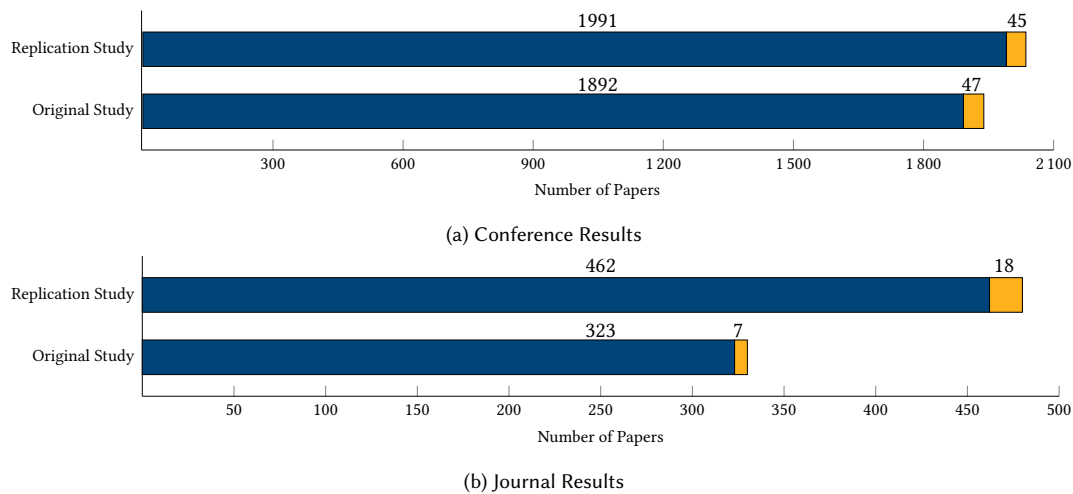
Naturally, we also compared our SLR results with the original SLR study \cite{hao2019ReplicationSLR} we were replicating. While the original SLR examined replication studies from 2009 to 2018, its findings covered only publications from 2011 to 2018, a period of eight years (since no replication studies  were found for the first two years). Their results bring the time frames for the two SLRs closer together, potentially minimising anomalies in the discussion of the results between the two studies. When comparing the overall replication studies identified in the two SLRs, the original SLR found 54 (2.38\%) replication studies, while ours found 63 (2.50\%), a comparable number in both.

Next, we compared results by conference/journal venue type, an approach used in the original SRL, previously presented in \autoref{section_related_work}. \autoref{fig:slr_paper_comparison} shows this comparison, with \autoref{fig:conference_stackbar} presenting the conference results, while \autoref{fig:journal_stackbar} shows the journal results. \autoref{fig:conference_stackbar} shows 47 (87\%) of the original SLR's replication papers published at conferences, while our SLR identified 45 (71.43\%). When comparing journal publications, the original SLR saw seven (13\%) replication studies, whereas our SLR identified 18 (29\%), demonstrating an increase in replication studies in journals.

\begin{figure}[t]
  \input{figures/context_stackedbar}
\caption{Comparing the Context Results Between Original and Replication Studies}\label{fig:slr_context_comparison}
\vspace{2mm}
\undergradbar{} Undergraduate \csonebar{} CS1 \elembar{} K-12 \gradbar{} Graduate \otherbar{} Other
\end{figure}
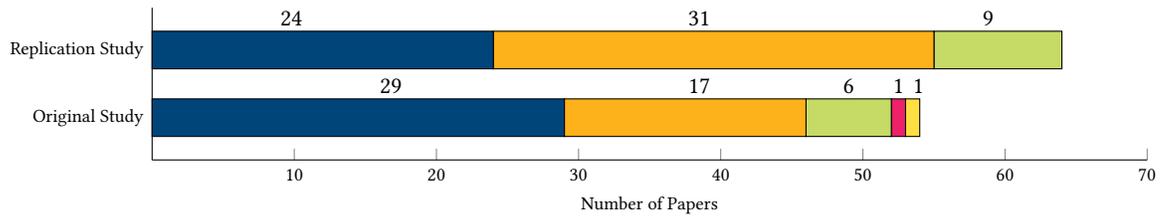

Then, we compared the Computing Education topics. The original SLR did not provide numeric information when quantifying results for topics. Instead, the authors presented the first five topics in the following order: \textit{Learning \& Teaching Strategies}, \textit{Assessment}, \textit{Learning Behaviour}, \textit{Learning Theory}, and \textit{Performance Prediction}. We also found that \textit{Learning \& Teaching Strategies} (N=17, 27\%) and \textit{Assessment} (N=14, 22\%) continue to be the most common topics in replication studies. However, our replication study found a higher level of interest evaluating \textit{Tools} (N=13, 20\%).


Next, we compared the contexts, as shown in \autoref{fig:slr_context_comparison}. The predominant context found in the original SLR was \textit{Undergraduate} (N=29, 54\%), followed by \textit{CS1} (N=17, 31\%), while \textit{Graduate} and \textit{Other} contexts had the least (N=1, 2\%) number of papers. Our SLR found that \textit{CS1} (N=31, 49\%) was the predominant context, followed by \textit{Undergraduate} (N=24, 38\%), whereas no replication studies used the \textit{Graduate} or \textit{Other} contexts.

\input{tables/comparison_replicationtypes}

\autoref{tab:comparing_slr_results} presents other comparisons between the two SLRs, examining methodologies, replication types, and authorship. 
For methodologies, \textit{Quantitative} remains the most common for both SLRs, with the original SLR study identifying 74\% (N=40) of their papers applying quantitative approaches, while we saw 60\% (N=38) of our papers utilising them. The least common approach across both SLRs was \textit{Qualitative}, with the original SLR identifying four papers (7\%) and our SLR identifying six (10\%). When comparing the replication types, our SLR identified authors applying different replication study designs, including \textit{Methodological} (N=8, 13\%) and \textit{Empirical} (N=4, 6\%) approaches. In contrast, the original SLR did not report on these methodologies. 

For authorship, the original SLR stated that ``the same authors, the replication success rate was 72\% and the failure rate was 22\%. In contrast, the success rate dropped to 58\% and failure rate increased to 28\% when all authors of a study were different'' \cite[p.~42]{hao2019ReplicationSLR}. Our results show that the same authors had a high success rate (82\%), with 18\% of the studies failing; by contrast, with different authors, the success rate dropped to 60\%, with the failure rate increasing to 25\%. We found that more of the same authors (N=35, 56\%) replicated their work than the original SLR found (N=18, 33\%).

Overall, our findings indicate a modest increase in the already limited prevalence of replication studies in CER compared to the original SLR, with a slight shift toward journal publications and a broader range of topics. The continued dominance of quantitative methodologies and the predominance of same-author replications suggest that, while the field is progressing, challenges remain in encouraging independent replications and diversifying methodological approaches. Notably, the context of replication studies has shifted, with CS1 now surpassing undergraduate contexts as the most common setting. These results highlight both areas of growth and persistent gaps, underscoring the need for ongoing efforts to promote replication across varied contexts, authorship models, and research designs within CER.

\begin{table}[b]
\caption{Publication Details by Survey Respondents Previously Conducting Replication Studies}
\label{tab:participants_replication_papers}
\begin{tabular}{l|ccc||cc}
\textbf{Question} & \textbf{N   } & \textbf{Mean} & \textbf{Median} & \textbf{Min} & \textbf{Max} \\
\hline
Q2. How many replication studies performed? & 29  & 2.66 & 2 & 1 & 9 \\
Q3. How many replication studies submitted for publication? & 29 & 1.97 & 1 & 0 & 9 \\
Q4. How many replication studies successfully published?\textsuperscript{*} & 27 & 1.74 & 1 & 0 & 8 \\
\hline
\multicolumn{6}{l}{* Two non-applicable responses} 
\end{tabular}
\end{table}
\subsection{Survey Results}\label{subsection_survey_results}


In the eight weeks during which the survey remained open, we recorded 116 responses. We asked all the participants \textit{``Have you implemented a published Computing Education Research (CER) intervention or work in your
classroom (i.e. done a replication study)?''} (Q1). We received \textit{``Yes''} from 53 (46\%) participants, while 63 (54\%) responded \textit{``No''}. 
We asked the 53 participants who answered \textit{``Yes``} additional questions including how many replication studies they had conducted (Q2), how many of those they had submitted for publication (Q3), and how many they had successfully published (Q4). \autoref{tab:participants_replication_papers} outlines their responses to these questions. For questions Q2-Q4, the table shows the number of responses, along with the mean, median, minimum, and maximum values provided by the participants. For Q4, two participants responded \textit{``not applicable''} because they answered \textit{``0''} to Q3. 
The answers showed that educators conducting replication studies are likely to submit most, if not all, of them for publication. Only five (17\%) of the 29 participants who reported performing several replications (Q2) answered that they submitted none of them for publication (Q3).

We asked participants whether they would voluntarily provide their previously published replication studies, which yielded 19 papers from 13 (11\%) participants. We found that 14 (74\%) came from the five venues we reviewed for the SLR. We received direct mention of six (32\%) papers from SIGCSE, four (21\%) from ITiCSE, two (11\%) from CSEJ, and two (11\%) from CSEJ. We found the remaining five (26\%) published at other venues: Koli Calling (N=2, 11\%), The Western Canadian Conference on Computing Education (WCCCE) (n=1, 5\%), Frontiers in Education (FIE) (N=1, 5\%), and The International Conference on the Future of Education, Teaching Systems and Information Technology (IFETS) (N=1, 5\%).

When evaluating the participants' 19 papers to determine whether they would meet our selection criteria, we found that one (5\%) paper appeared in the original SLR \cite{hao2019ReplicationSLR}, while nine (47\%) appeared in our SLR. We observed that four (21\%) would not have met our search criteria because the term \textit{``replicat[a-z]*''} did not appear in the publications. Two (11\%) papers would also have been excluded under our selection criteria because they did not explicitly state that the studies performed replication, while the remaining three (16\%) were not yet published, so we could not evaluate them.

Of the 29 participants with experience in replication, seven (24\%) reported that venues rejected their replication papers during the peer-review process, identifying SIGCSE, ITiCSE, ICER, Koli Calling, and ``ACM and Taylor \& Francis'' as the submission venues. When we asked participants whether they would consider conducting replication studies in the future, 56\% were unsure, 34\% said \textit{``Yes''}, and 30\% said \textit{``No''}.



\begin{table}[]
    \caption{Qualitative Coding Scheme for Perceived Value of Replication (Q8) and Explanation of Limited Replication in CER (Q10). (Percentages of unique respondents discussing the theme do not sum up to 100\% because responses could contain multiple themes. Total number of open-text responses N = 147)}
    \label{tab:Q8+Q10-themes-A}
    \centering
    \begin{tabularx}{\textwidth}{p{3cm}|p{3cm}|c|B}
    \hline
    \makecell[l]{\textbf{Value / Barrier}\\\textbf{Dimension}} & 
    \makecell[l]{\textbf{Associated Thematic}\\\textbf{Tags}} &
    \makecell[c]{\textbf{Resp}\\\textbf{(\%)}} & 
    \makecell[l]{\textbf{Theme Summary}} \\
    \hline
        
        
        \multirow{8}{*}{\parbox{2.8cm}{\raggedright Building Trustworthy Knowledge}} 
            & Generalisability & 37\% & 
            \parbox{6.7cm}{\vspace{2pt}(value, barrier) Replication tests if findings hold across different contexts, populations, or institutions\vspace{2pt}}\\
            \cline{2-4} 
            & Confirmation of Knowledge & 37\% & 
            \parbox{6.7cm}{\vspace{2pt}(value) Replication validates, confirms, or strengthens confidence in existing results \vspace{2pt}}\\
            \cline{2-4}
            & Knowledge Gaps & 24\% & 
            \parbox{6.7cm}{\vspace{2pt} (value) Replication is a way to identify limitations, missing evidence, or areas where current knowledge is insufficient; (barrier) Replication is necessary but hindered because existing studies lack sufficient methodological detail or accessible materials to enable replication\vspace{2pt}}\\
            \cline{2-4}
            & Contradictory Results & 12\% & 
            \parbox{6.7cm}{\vspace{2pt}(value, barrier) Replication can reveal inconsistencies or challenge prior findings\vspace{2pt}}\\ 
        \hline

        \multirow{7}{*}{\parbox{2.4cm}{\raggedright Supporting\\the Field}} 
            & Valuable & 74\% & 
            \parbox{6.7cm}{\vspace{2pt}(value) Replication is important, worthwhile, or necessary to CER\vspace{2pt}}\\
            \cline{2-4} 
            & Beneficial to the Field & 16\% & 
            \parbox{6.7cm}{\vspace{2pt}(value) Replication contributes to the health, credibility, or development of the CER community or discipline, (barrier) even though it is structurally unsupported, highlighting the disconnect between epistemic importance and institutional reward\vspace{2pt}}\\
            \cline{2-4}
            & Underutilised & 14\% & 
            \parbox{6.7cm}{\vspace{2pt}(barrier) Replication is insufficiently used, rare, or neglected within CER\vspace{2pt}}\\
        \hline

            \multirow{9}{*}{\parbox{2.4cm}{\raggedright Incentives, Recognition, Reward Structures and Publication Culture}} 
            & Novelty & 58\% & 
            \parbox{6.7cm}{\vspace{2pt} (barrier) The dominance of novelty, originality, or ``newness'' is a cultural or review expectation that disadvantages replication\vspace{2pt}}\\
            \cline{2-4}
            & Research Recognition & 24\% & 
            \parbox{6.7cm}{\vspace{2pt}(barrier) Replication receives limited recognition within the research community (e.g., in publication prestige, citations, or scholarly status)\vspace{2pt}}\\
            \cline{2-4}
            & Professional Recognition & 20\% & 
            \parbox{6.7cm}{\vspace{2pt}(barrier) Replication is undervalued in career-related contexts (e.g., hiring, tenure, promotion, performance evaluation)\vspace{2pt}}\\ 
            \cline{2-4} 
            & Undervalued & 15\% & 
            \parbox{6.7cm}{\vspace{2pt}(barrier) Replication is culturally perceived as less important or prestigious\vspace{2pt}}\\
            \cline{2-4}
            & Reviewer Bias & 15\% & 
            \parbox{6.7cm}{\vspace{2pt}(barrier) Peer-review practices or reviewer expectations disadvantage non-novel work\vspace{2pt}}\\
        \hline
    \end{tabularx}
\end{table}

We asked all participants two open-ended questions: \textit{``What are your opinions on the value of replication studies in CER?''} (Q8) and \textit{``Why do you think the CER community does not do more replication studies?''} (Q10). The responses frame seven overarching value dimensions, with corresponding themes presented next and outlined in \autoref{tab:Q8+Q10-themes-A} and \autoref{tab:Q8+Q10-themes-B}.

\subsubsection{High Epistemic Value and Generalisability}

Our survey participants consistently framed open-text responses on replication as central to the development of trustworthy knowledge in CER. Rather than viewing replication as redundant, they emphasised its role in determining if findings extend beyond a single study or a specific instructional context. As one participant noted, \textit{``High value: it's good to know whether an intervention done in one context is more broadly effective'' (R003)}. Others echoed this concern with transferability and context, writing, `\textit{`Valuable - especially in sufficiently different contexts that they increase our belief in the general validity of the idea'' (R005)} and \textit{``They are valuable. Context is a significant issue in CER studies, which only replication can finally address'' (R015)}. 

\begin{table}[]
    \caption{Qualitative Coding Scheme for Perceived Value of Replication (Q8) and Explanation of Limited Replication in CER (Q10). (Percentages of unique respondents discussing the theme do not sum up to 100\% because responses could contain multiple themes. Total number of open-text responses N = 147)}
    \label{tab:Q8+Q10-themes-B}
    \centering
    \begin{tabularx}{\textwidth}{p{3cm}|p{3cm}|c|B}
    \hline
    \makecell[l]{\textbf{Value / Barrier}\\\textbf{Dimension}} & 
    \makecell[l]{\textbf{Associated Thematic}\\\textbf{Tags}} &
    \makecell[c]{\textbf{Resp}\\\textbf{(\%)}} & 
    \makecell[l]{\textbf{Theme Summary}} \\
    \hline

            \multirow{4}{*}{\parbox{2.8cm}{\raggedright Practical Challenges}} &
            Hard to Do & 30\% & 
            \parbox{7.2cm}{\vspace{2pt}(barrier) Replication is methodologically, logistically, or practically difficult to carry out\vspace{2pt}}\\
            \cline{2-4}
            & Hard to Publish & 18\% & 
            \parbox{7.2cm}{\vspace{2pt}(barrier) Replication is difficult to place in appropriate venues or to have accepted for publication\vspace{2pt}} \\
            \cline{2-4} 
            & Hard to Write & 3\% &
            \parbox{7.2cm}{\vspace{2pt}(barrier) Replication is difficult to frame, report, or argue for in a way that fits publication norms\vspace{2pt}}\\
        \hline

            \multirow{3}{*}{\parbox{2.8cm}{\raggedright Context and Ethical Constraints}} &
            Context & 45\% & 
            \parbox{7.2cm}{\vspace{2pt}(barrier) Replication influenced by contextual factors (e.g., institutional, cultural, curricular differences) that complicate replication or comparability\vspace{2pt}}\\
            \cline{2-4} 
            & Ethics & 4\% & 
            \parbox{7.2cm}{\vspace{2pt}(barrier) Ethical constraints (e.g., consent, student risk, intervention withholding) limit replication design or feasibility\vspace{2pt}}\\
        \hline

            \multirow{1}{*}{\parbox{2.8cm}{\raggedright Definitional Ambiguity}} & 
            Replication Adjacent & 8\% & 
            \parbox{7.2cm}{\vspace{2pt}(barrier) Not a ''pure'' replication but blends confirmation with extension, adaptation, or contextual variation\vspace{8pt}}\\
            
        \hline

            \multirow{1}{*}{\parbox{2.8cm}{\raggedright Lack of Interest}} &
            Not Interested & 8\% & 
            \parbox{7.2cm}{\vspace{2pt}(barrier) Lack of personal or community interest in conducting replication studies\vspace{2pt}}\\
        \hline
    \end{tabularx}
\end{table}

\subsubsection{Confirmation of Knowledge, Exposing Knowledge Gaps}
We also report that participants repeatedly described replication as a mechanism for strengthening confidence in existing results. One participant wrote, \textit{``Would be good to do more. Would let us confirm the validity of findings'' (R070)}, while another added, \textit{``I think they are useful as they can bring more credibility to previous findings'' (R076)}. In this sense, R076's response positions replication as part of building a cumulative evidence base rather than producing isolated findings, as reflected in the comment, \textit{``They're critical for establishing computing education as science/social science. We have to be able to test and build on prior work'' (R010).}

Importantly, participants not only framed replication as confirmatory but also emphasised its value in exposing limitations and gaps in the literature. One noted, \textit{``[Replication studies] are crucial to improving our collective understanding'' (R024)}, and another observed, \textit{``I would encourage more of them! I think a tendency toward novelty above all else---except in circumstances where we are looking into new things and exploration is perhaps more valuable than certainty---can leave gaps in our knowledge. Particularly since generalisation is difficult in contexts as population dependent as education, significant amounts of replication can lead to a deeper understanding of how these findings vary across populations'' (R080)}.

Other survey participants were more explicit about the role of replication in surfacing contradictions. For example, R034 stated, \textit{``They are incredibly valuable, but the nature of replication sits slightly at odds with the innate variation of participants found across cultural and socio-temporal boundaries - there is a question as to exactly how much we can replicate. [...] Replication studies should be carried out on all CER research as a matter of course, given how many studies we are already aware of that cannot be replicated'' (R034)}.

\subsubsection{Beneficial to the Field but Underutilised and Undervalued}
Beyond its epistemic role, participants widely described replication as work that supports the CER field as a whole. Participants expressed this concisely and forcefully, stating \textit{``Tremendous value'' (R022)}, \textit{``Must be done'' (R021)}, and \textit{``Urgently needed and hugely underappreciated'' (R074)}. These statements reflect a perception of replication as foundational to the credibility and health of CER, even when it does not directly benefit individual researchers. Participants highly valued replication in CER, making its value the most-discussed topic, with nearly three-quarters commenting on it. One succinctly described the underutilisation of replication, stating, \textit{``There are so few replication studies that I wonder whether we can actually call ourselves a ``science'''' (R074)}.

\subsubsection{The Expectation of Novelty, Definitional Ambiguity and Reviewer Bias}
Some participants pointed to ambiguity around what should count as replication, particularly from a reviewer and publication standpoint. For example, R068 stated, \textit{``I also received feedback on a rejected paper from a reviewer that gave the impression they saw the replication study needing to be exact replication, which my wasn't. Perhaps rejected replication papers due to misunderstandings on the types of replication studies available may be why the community doesn't do more of these studies'' (R068)}. Participants mentioned that partial replication and replication with extensions are also hard to publish.

Participants repeatedly described replication as underutilised and insufficiently rewarded. One participant summarised this by describing it as \textit{``Valuable, but not rewarded as easily as original research'' (R075)}. Another participant stated, \textit{``I think they're super important, but also hard to publish because reviewers tend to say ``what's the novelty in this?'''' (R004)}.

Survey participants noted a significant mismatch between the scientific importance of replication and the ways CER evaluates and rewards it. Participants explicitly linked this to novelty-driven review cultures, with one noting, \textit{``There's too much of a premium on ``novelty'' in CS research in general. Replication studies almost always get viewed as less interesting because of the perception of lack of novelty'' (R003)}. Another wrote, \textit{``Reviewers negatively comment on the `value' and `novelty' of the work'' (R029)}, and a third explained, \textit{``They're hard to publish. I think they're looked down on as not being novel or rigorous'' (R004)}. As a matter of fact, the novelty bias issue was the second-most-often addressed topic in participants' written responses, with more than half discussing it. 

Responses also mentioned reviewer bias as a vexation in connection with study setting, \textit{``frustrations over the review process; For example, some reviewers might complain that the student populations in each study are ``too different'' when in fact, that the entire point of doing the replication study'' (R098)}. Participants raised issues with CER venues, for example, \textit{``SIGCSE/ICER/ITICSE are the only real outlets for CER, and they don't put much value on replication studies'' (R086)}.

Participants not only mentioned novelty related to reviewer bias, but also noted that it is a reason the CER community does not conduct more replication studies. One train of thought was that such work will not be valuable for job seekers, \textit{``Since much of research is driven by graduate students who will have to peacock in front of search committees to get jobs, being able to demonstrate your own intelligence is easier to do when you're doing ``novel'' work rather than replicating'' (R080)}. Others pointed to the CER community culture as the culprit. For example, participant R002 stated, \textit{``An oft-cited reason is that they are not valued. But I think CER does value them, and people are usually happy to see them. I think it's more that people are unexcited by the idea of taking someone else's research design and re-running it; they want the fun and invention and novelty of doing their own design'' (R002)}. Another participant stated, \textit{``Likely more interested in chasing the next shiny object and not having to argue about why it is needed when it has been done before'' (R104)}, while another participant wrote, \textit{``I believe that scientists find replication studies as less-than. They believe that this is not an original study, therefore not as important as an original idea/intervention. Unfortunately, this hurts science!'' (R037)}.

\subsubsection{Lack of Incentives and Recognition}
The responses emphasised novelty tied to broader existing incentive structures, such as the common topics of research and professional recognition, or the lack thereof. As one participant stated, \textit{``No motivation, equal work with low probability of publication, and low citation count if it is published'' (R036)}. Participant R072 noted replication studies receive less recognition than the original, novel study, stating, \textit{``There is a sense that less ``credit'' is given to replications. If the replication study confirms the original study, then in some sense the credit goes to the original study designer'' (R072)}. Another participant summarised the structural misalignment as \textit{``Lack of incentive structure, grant funding not aligned'' (R033)}, with participant R025 explaining, \textit{``for those who need tenure, replication studies may not provide the value that tenure and promotion committees are looking for'' (R025)}. In this context, survey participants described replication as valuable and legitimate work while also emphasising the lower professional payoff, such as one participant stating, \textit{``It is just a lot of work and very unlikely to pay off for a researcher'' (R098)}.

\subsubsection{Practical Challenges and Ethical Considerations}
Participants raised the practical difficulty of conducting replication. One participant wrote, \textit{``[Replication studies] would be very useful, but very challenging due to local constraints and educational setups'' (R007)}. Others addressed difficulties in reporting replication, such as one participant stating, \textit{``It also seems harder to write a replication study paper'' (R033)} and another stating, \textit{``Certainly worthwhile, but I have never written or reviewed one. I would be nervous about submitting one and am unsure how I would structure such a paper'' (R023)}.

Participants also raised concerns with replication studies requiring IRB approval, acknowledging it as a constraint on replication. For example, one participant stated, \textit{``I don't know about other researchers, but a major barrier for me is getting the necessary ethics approval'' (R021)}. Another participant mentioned the overhead of obtaining IRB approval, stating, \textit{``getting IRB approval and setting up the whole thing correctly takes a lot of time and coordination'' (R099)}.

\subsubsection{Context}
Context was commented on extensively by 40\% of our survey participants. They emphasised that the educational context fundamentally shapes what replication means in CER, and that differences across institutions, students, and instructional settings complicate both the design of replication studies and the interpretation of their findings, with generalisability as the central issue. For example, one participant stated, \textit{``It can be an opportunity to make sure the approach is still relevant and/or extend findings to a new context'' (R104)}. Participants posit that context makes replication necessary, and that this is a core challenge in conducting replication, not a nuisance variable that researchers can control away. Additionally, the lack of information and context provided by original studies also compounds the complexity for conducting replication, with participant R007 stating, \textit{``Publications seldom include enough information about the context, making difficult to evaluate possible replication of a study'' (R007)}. 


Despite these practical challenges, one participant perceives replication occurring in CER, but remaining largely invisible, stating \textit{``I think [the CER community does replication studies], they just don't publish them'' (R027)}.

\subsubsection{The Risk of Contradictory Results}
Finally, a recurrent concern among our participants related to the consequences of producing contradictory results, especially in seminal studies. For example, one participant stated, \textit{``concerns of a null result being at odds with accepted fact'' (R034)}.

Participants noted that replication is often implicitly expected to confirm prior work, and that studies which fail to reproduce original findings can be particularly difficult to position and publish. Participants described a perceived asymmetry in how confirmatory versus contradictory replications are received. For example, one participant stated, \textit{``People may worry what happens if they get contradictory results; is this viewed as an insult to the original authors?'' (R002)}. Another noted, \textit{``If people publish replication studies and the results are negative. Often, the community attacks the researchers. I have seen this in other communities outside of CER'' (R095)}. Another participant stated, \textit{``I hear from folks it can be awkward because a person might discover faults in the original study'' (R091)}, and participant R072 explained, \textit{``If the replication study does not confirm the original study, then there is a bit of an implied hostility that could be tricky to manage'' (R072)}. These responses related to potential hostilities are worrisome for some participants, with one noting that \textit{``if the results do not replicate (lack verification or validation) someone could fear making an enemy in the community when publishing'' (R104)}. Participant's R104 response reflects perceived cultural risks associated with challenging established findings and helps explain why some replications may never be submitted or published. 

\subsubsection{What Should We Do?}
The final survey question asked participants whether they had anything to add before completing the survey. Some of the parting thoughts suggested that incentives, such as funding, awards, and smaller dedicated events, could encourage more replication in CER, along with introducing special workshops and venues that designate special issues and special tracks for replication. For example, one participant stated, \textit{``If the dearth of replication studies is, as I believe, a system issue related to incentives, then it needs a systemic, cultural approach to solving it. Create dedicated tracks within top conferences, hold workshops and discussion sessions, publish a whitepaper with broad community authorship and support about the importance and value of replication studies, etc.'' (R006)}.

Participants also addressed the need to strengthen the permissibility of contradictory results, such as one participant stating, \textit{``ACM should have specific tracks for replications (and null result experiments) with lower barrier to entry, this could make it easy for newcomers, or graduate students to get some ``starter publications'''' (R036)}.

We also received responses emphasising the need for reviewer training. For example, \textit{``I think we need to do a lot more of them! And reviewers need to better understand the value of doing them'' (R029)}. One participant summarised \textit{``It would be interesting to consider a separate review track for it; There is usually a question on novelty of the work, and it seems like some reviewers get hung up on that. Perhaps a separate track or separate guidance on replication studies would help reviewers better navigate; I really hope we see more of them!'' (R098)}.

We observed training arguments interleaved with the need for sufficient context sharing. For example, \textit{``We should be writing up our research so everything can be replicated, sharing our data sets, but building a review and replication analysis community that supports positive and constructive reviews to develop research skills and improve our knowledge. There is a very big difference between constructive and honest feedback and ad hominen / ex authoritas arguments to crush failed replications of prominent studies'' (R034)}. Participants emphasised the need to request studies to share details to enable replication, such as \textit{``Maybe if the journals and conferences somehow require open-source code and datasets availability, we are going to see more and more studies on the replication studies in CER'' (R094)}, and \textit{``Having explicit paper tracks for replication would be helpful, also, making explicit when reviewing papers, that enough detail should be provided so someone else can replicate their study'' (R113)}.

Participants also wrote that the culture of the CER community needs to change for replication to become more prevalent. For example, one participant stated, \textit{``Our community just needs to mature in attitudes. But, most of us are PhDs and we believe (trained to be?) inventors and experimenters. We are not trained to adopt others' work'' (R082)}.

Overall, our survey responses show that the challenge for CER is not whether the community values replication, but whether current research cultures and incentive structures make it feasible, visible, and professionally sustainable.

%% file: tables/slr_yearly_results.tex
\begin{table}[t]
\caption{SLR Results Across CER Venues, Including Total (Tot) and Replication Studies (Rep) Found in Each Venue and Year}
\label{tab:slr_results}
\small
\begin{tabular}{l||ll|ll||ll|ll|ll||ll}
&\multicolumn{10}{c||}{\textbf{Venues}}&\\
\cline{2-11}
&\multicolumn{4}{c||}{\textbf{Journals}}&\multicolumn{6}{c||}{\textbf{Conferences}}&\\
\textbf{Year} & \multicolumn{2}{c|}{\textbf{CSEJ}} & \multicolumn{2}{c||}{\textbf{TOCE}} & \multicolumn{2}{c|}{\textbf{ICER}} & \multicolumn{2}{c|}{\textbf{ITiCSE}} & \multicolumn{2}{c||}{\textbf{SIGCSE}} & \multicolumn{2}{c}{\textbf{Total}} \\
\hline
\rowcolor{lightgray} 
& Tot & Rep & Tot & Rep & Tot & Rep & Tot & Rep & Tot & Rep & Tot & Rep\\
2019 & 15 & 1 (7\%) & 40 & 1 (3\%) & 28 & 1 (4\%) & 66 & 0 (0\%)\textsuperscript{\ddagger} & 179 & 5 (3\%) & 328 & 8 (2\%) \\
2020 & 18 & 0 (0\%)\textsuperscript{\ddagger} & 27 & 0 (0\%)\textsuperscript{\ddagger} & 27 & 1 (4\%) & 72 & 3 (4\%) & 171 & 3 (2\%) & 315 & 7 (2\%) \\
2021 & 18 & 0 (0\%)\textsuperscript{\ddagger} & 46 & 0 (0\%)\textsuperscript{\ddagger} & 30 & 2 (7\%) & 84 & 3 (4\%) & 170 & 2 (1\%) & 348 & 7 (2\%) \\
2022 & 19 & 4 (21\%)\textsuperscript{\dagger} & 43 & 3 (7\%) & 25 & 0 (0\%)\textsuperscript{\ddagger} & 79 & 1 (1\%) & 144 & 2 (1\%) & 310 & 10 (3\%) \\
2023 & 24 & 0 (0\%)\textsuperscript{\ddagger} & 30 & 1 (3\%) & 35 & 1 (3\%) & 80 & 5 (6\%) & 165 & 3 (2\%) & 334 & 10 (3\%) \\
2024 & 32 & 1 (3\%) & 57 & 1 (2\%) & 36 & 5 (14\%) & 107 & 2 (2\%) & 216 & 1 (1\%) & 448 & 10 (2\%) \\
2025 & 21 & 0 (0\%)\textsuperscript{\ddagger} & 90 & 6 (7\%) & 31 & 1 (3\%) & 99 & 0 (0\%)\textsuperscript{\ddagger} & 192 & 4 (2\%) & 433 & 11 (3\%) \\
\hline\hline
\multirow{2}{*}{\textbf{Total}} & 147 & 6 (4\%) & 333 & 12 (4\%) & 212 & 11 (5\%) & 587 & 14 (2\%) & 1237 & 20 (2\%) & \multirow{2}{*}{2516} & \multirow{2}{*}{63 (3\%)}\\
& \multicolumn{4}{c||}{\textbf{Journal:} Tot=480  Rep=18 (4\%)} &\multicolumn{6}{c||}{\textbf{Conference:} Tot=2036 Rep=45 (2\%)}\\
\multicolumn{13}{l}{\newline \dagger Includes special issue on replication studies\hspace*{10mm} \ddagger Year for venue with no reported replication studies}
\end{tabular}
\end{table}

%% file: figures/conference_stackedbar.tex
\begin{tikzpicture}
\begin{axis}[
    xbar stacked,
    clip=false, 
    width=.98\textwidth,
    bar width=5mm,
    enlarge y limits={abs=0.625},
    ytick=data,
    xmin=-3,
    xmax=2100,
    ytick style={draw=none},
    axis y line*=none,
    axis x line*=bottom,
    xlabel = {Number of Papers},
    tick label style={font=\footnotesize},
    label style={font=\footnotesize},
    xtick={300,600,900,1200,1500,1800,2100},
    yticklabels={Original Study, Replication Study},
    area legend,
    visualization depends on=x \as \XVal,
    y=9mm,
    /pgf/number format/1000 sep=\,
]
\addplot[fill=nonreplication] coordinates{(1892,0) (1991,1)};
\addplot[fill=replication] coordinates{(47,0) (45,1)};

\node[text=black, anchor=center] at (rel axis cs:0.5, 0.47) {1892};
\node[text=black, anchor=center] at (rel axis cs:0.91, 0.47) {47};

\node[text=black, anchor=center] at (rel axis cs:0.5, 0.93) {1991};
\node[text=black, anchor=center] at (rel axis cs:0.96, 0.93) {45};

\end{axis}  
\end{tikzpicture}

%% file: figures/journal_stackedbar.tex
\begin{tikzpicture}
\begin{axis}[
    xbar stacked,
    clip=false,
    width=.98\textwidth,
    bar width=5mm,
    enlarge y limits={abs=0.625},
    ytick=data,
    xmin=0,
    xmax=500,
    ytick style={draw=none},
    axis y line*=none,
    axis x line*=bottom,
    xlabel = {Number of Papers},
    tick label style={font=\footnotesize},
    label style={font=\footnotesize},
    xtick={50,100,150,200,250,300,350,400,450,500},
    yticklabels={Original Study, Replication Study},
    area legend,
    visualization depends on=x \as \XVal,
    y=9mm,
    /pgf/number format/1000 sep=\,
]
\addplot[fill=nonreplication] coordinates{(323,0) (462,1)};
\addplot[fill=replication] coordinates{(7,0) (18,1)};

\node[text=black, anchor=center] at (rel axis cs:0.5, 0.47) {323};
\node[text=black, anchor=center] at (rel axis cs:0.65, 0.47) {7};

\node[text=black, anchor=center] at (rel axis cs:0.5, 0.93) {462};
\node[text=black, anchor=center] at (rel axis cs:0.94, 0.93) {18};

\end{axis}  
\end{tikzpicture}

%% file: figures/context_stackedbar.tex
\begin{tikzpicture}
\begin{axis}[
    xbar stacked,
    clip=false,
    width=.98\textwidth,
    bar width=5mm,
    enlarge y limits={abs=0.625},
    ytick=data,
    xmin=0,
    xmax=70,
    ytick style={draw=none},
    axis y line*=none,
    axis x line*=bottom,
    xlabel = {Number of Papers},
    tick label style={font=\footnotesize},
    label style={font=\footnotesize},
    xtick={10,20,30,40,50,60,70},
    yticklabels={Original Study, Replication Study},
    area legend,
    visualization depends on=x \as \XVal,
    y=9mm,
    /pgf/number format/1000 sep=\,
]
\addplot[fill=nonreplication] coordinates{(29,0) (24,1)};
\addplot[fill=replication] coordinates{(17,0) (31,1)};
\addplot[fill=SpringGreen] coordinates{(6,0) (9,1)};
\addplot[fill=WildStrawberry] coordinates{(1,0) (0,1)};
\addplot[fill=Goldenrod] coordinates{(1,0) (0,1)};

\node[text=black, anchor=center] at (rel axis cs:0.14, 0.93) {24};
\node[text=black, anchor=center] at (rel axis cs:0.55, 0.93) {31};
\node[text=black, anchor=center] at (rel axis cs:0.84, 0.93) {9};

\node[text=black, anchor=center] at (rel axis cs:0.24, 0.49) {29};
\node[text=black, anchor=center] at (rel axis cs:0.55, 0.49) {17};
\node[text=black, anchor=center] at (rel axis cs:0.70, 0.49) {6};
\node[text=black, anchor=center] at (rel axis cs:0.75, 0.49) {1};
\node[text=black, anchor=center] at (rel axis cs:0.77, 0.49) {1};
\end{axis}  
\end{tikzpicture}

%% file: tables/comparison_replicationtypes.tex
\begin{table}[b]
\caption{Original (Org) and Replication (Rep) SLR Study Results by Methodology, Replication (Rep) Type, and Authorship\\(A dash (-) represents category not reported)}
\label{tab:comparing_slr_results}
\small
\begin{tabularx}{\textwidth}{X|cc|cc|cc|cc|cc}
\hline
\rowcolor{lightgray} 
\textbf{Methodology} & \multicolumn{2}{c|}{\textbf{Total}} & \multicolumn{2}{c|}{\textbf{Success}} & \multicolumn{2}{c|}{\textbf{Mixed}} & \multicolumn{2}{c|}{\textbf{Failure}} & \multicolumn{2}{c}{\textbf{Unknown}} \\
\hline
& Org & Rep & Org & Rep & Org & Rep & Org & Rep & Org & Rep \\ 
\cline{2-11} 
Quantitative & 40 (74\%) & 38 (60\%) & 27 (68\%) & 29 (76\%) & 9 (22\%) & 1 (3\%)  & 4 (10\%) & 8 (21\%) & - & 0 (0\%) \\
Mixed & 10 (19\%)& 19 (30\%) & 5 (50\%) & 12 (63\%) & 5 (50\%) & 3 (15\%) & 0 (0\%) & 2 (11\%) & - & 2 (11\%) \\
Qualitative & 4 (7\%) & 6 (10\%) & 2 (50\%) & 4 (66\%) & 2 (50\%) & 0 (0\%) & 0 (0\%) & 1 (17\%) & - & 1 (17\%) \\
\hline
\rowcolor{lightgray} 
\textbf{Rep Type} & \multicolumn{2}{c|}{\textbf{Total}} & \multicolumn{2}{c|}{\textbf{Success}} & \multicolumn{2}{c|}{\textbf{Mixed}} & \multicolumn{2}{c|}{\textbf{Failure}} & \multicolumn{2}{c}{\textbf{Unknown}} \\
\hline
& Org & Rep & Org & Rep & Org & Rep & Org & Rep & Org & Rep \\ 
\cline{2-11}
Conceptual & 41 (76\%) & 41 (65\%) & 27 (66\%) & 28 (68\%) & 3 (7\%) & 3 (7\%)& 11 (27\%) & 7 (18\%) & - & 3 (7\%)\\
Direct & 13 (24\%) & 10 (16\%) & 7 (54\%) & 8 (80\%) & 3 (23\%) & 0 (0\%) & 3 (23\%) & 2 (20\%) & - & 0 (0\%) \\
Methodological & - & 8 (13\%) & - & 6 (75\%) & - & 0 (0\%) & - & 1 (12.5\%) & - & 1 (12.5\%) \\
Empirical & - & 4 (6\%) & - & 4 (100\%) & - & 0 (0\%) & - & 0 (0\%) & - & 0 (0\%) \\
\hline 
\rowcolor{lightgray} 
\textbf{Authorship} & \multicolumn{2}{c|}{\textbf{Total}} & \multicolumn{2}{c|}{\textbf{Success}} & \multicolumn{2}{c|}{\textbf{Mixed}} & \multicolumn{2}{c|}{\textbf{Failure}} & \multicolumn{2}{c}{\textbf{Unknown}} \\
\hline
& Org & Rep & Org & Rep & Org & Rep & Org & Rep & Org & Rep \\ 
\cline{2-11}
Same  & 18 (33\%)& 35 (56\%) & 13 (72\%) & 29 (82\%) & 1 (6\%) & 0 (0\%) & 4 (22\%) & 3 (18\%) & - & 3 (18\%) \\
Different  & 36 (67\%)& 28 (44\%) & 21 (58\%) & 17 (60\%) & 5 (14\%) & 3 (11\%) & 10 (28\%) & 7 (25\%) & - & 1 (4\%)\\
\hline
\end{tabularx}
\end{table}

%% file: 06discussion.tex
\section{Discussion}\label{section_discussion}





\subsection{RQ1: How has the adoption of replication study design changed over the last seven years?}

The most striking aspect of our findings is that while replication persists, its growth has not matched the expectations set by recent discussions on open science and research rigour~\cite{Korbmacher2023}. We found that the overall proportion of published replication studies remains small and remarkably stable, with more researchers replicating their work and publishing in journals. Though the number of published replications remain small, our survey found that many researchers actively engage in replication in their own teaching and research, even if that work never reaches publication. Taken together, this suggests that adoption has increased in practice more than in print. 

We found that replication occurs in CER, but it is not consistently visible. A potential reason may be researchers' apprehension about publishing findings that do not align with the original findings. Our survey participants expressed particular concern about the consequences of producing contradictory results, including increased scrutiny, defensive reviewing, and professional risk when challenging well-established study findings. Replications that fail to confirm earlier work are less visible in peer-reviewed publications, resulting in literature biased towards positive confirmations. Future work could build on our findings and explore how researchers' apprehensions influence their willingness to publish or pursue replication studies that could yield contradictory results.    

We also found the CER community supports replication as a legitimate design, yet the structures that determine what gets published have not changed over time. This support may have influenced the growth in replication in CER publications, and the CER community has recently promoted replication. For example, \citet{brown2022launching} suggested a future journal initiative that includes an additional review cycle with an initial stage of registered replication reports. These efforts could encourage more replication, and the CER community could introduce such initiatives to conferences. 



\subsection{RQ2: What perceptions do Computing Education researchers have of using replication in their work?}
We found little disagreement about the value of replication across our survey responses, in which participants consistently described it as important for verifying knowledge, strengthening the field, and understanding the limits of existing results. Our participants framed replication as part of responsible research practice. We found 147 (6\%) papers during our SLR that encouraged replication of their work, and observed authors providing instruments, treatments, interventions, and datasets, where possible, for replication. From prior efforts \cite{hao2019ReplicationSLR}, researchers, such as \citet{poulsen2022insightsSQLHWProbs}, were inspired to conduct a replication study.  

At the same time, our survey participants' positive views of replication were paired with clear reservations. They consistently described replication works as high-effort, high-risk, and low-reward. Participants strongly endorsed the value of replication, but also reported misalignment with prevailing incentive structures, particularly those that prioritise novelty and rapid publication. In 2016, \citet{guzdial2016Miranda} observed a lack of support for replication that may persist today during the peer-review process. This suggests future work to determine whether CER venues and evaluation systems are set up in ways that actually make replication feasible and worthwhile.

Survey participants perceived replication as demanding, time-consuming, and risky, especially when results did not align with the original study. However, nearly half of the survey responses reported having implemented at least one published CER intervention in their own teaching or research, suggesting substantial replication beyond peer-reviewed publications. This could explain why the proportion of published replications has remained stable despite the increasing community attention and support on the topic. We found participants perceived that replication is harder to write, harder to publish, and less likely to be rewarded than novel work. In practice, this creates a chasm between what researchers believe is important to do and what they feel is professionally sustainable and advantageous.

Overall, we found that the CER community respects the importance of replication, but the effort and risks are not easy to justify. In this sense, the SLR's stable replication rate does not reflect a lack of interest or engagement, but a filtering effect: only a small subset of replication studies are perceived as sufficiently publishable under current norms. However, as more researchers replicate their work, some may not view it as replication, which future work could help clarify the criteria and distinctions for replication studies \cite{sigsoftReplicationStandard} to strengthen the CER community's understanding and practice of this research design. Our survey responses show that very few view their work as a ``pure'' replication, potentially demonstrating the community's perception that replication is ``exact'', where studies attempt to reproduce the original study's design and procedures under the same conditions \cite{huffmeier2016ReplicationTypology}. Instead, most participants describe replication as some form of adaptation or extension, shaped by local context and practical constraints. 





\subsection{RQ3: How can the CER community support researchers in adopting the replication study design?}


Overall, both the SLR and the survey support structural changes in our venues and evaluation systems to increase the number of publications that apply replication in CER. Fortunately, we have noticed positive changes over the last seven years. Firstly, we observed that venues matter for publishing replication studies. Journal initiatives and special issues for replication have clearly made a difference. 



Secondly, our results show that review practices need to change. Participants' concerns about novelty bias and the treatment of contradictory results suggest that replication must be evaluated on rigour and transparency, not on whether it confirms or extends prior work. In addition, our survey results highlight why more replication studies are appearing in journals rather than conferences. Participants emphasised the need for space to justify context, document methodological variation, and provide supporting materials. Our findings strengthen previous work by \citet{ihantola2015DataMiningLitReview}, who found that page-length restrictions in conference proceedings often limit the details available for replication in CER. From this perspective, publishing more in journals is not simply a change in venue preference, but a response to the reporting demands that credible replication in CER entails.   

Thirdly, better support for sharing materials and methods would lower the barrier to entry. Many respondents emphasised the difficulty of accessing sufficient detail to replicate studies well. Normalising artefact sharing and reuse would directly support higher-quality replication. Fortunately, services such as figshare and Center for Open Science (COS) provide platforms for sharing research materials and promoting transparency in instruments, research materials, and data and can be used by both journal and conference publications.\footnote{figshare: https://figshare.com/; COS: https://www.cos.io/products/osf} While some data, such as those requiring ethics (IRB) board approval, may continue to pose challenges for replication, this presents a valuable opportunity for future research to explore ways of providing data that meet IRB requirements.

Finally, for our survey participants, recognition matters. If replication continues to be undervalued in hiring, promotion, and evaluation, it will remain a secondary activity. Treating replication as a legitimate scholarly contribution rather than lesser work is essential for sustained adoption. When the CER community explicitly welcomes replication, and the review criteria reflect this, researchers are more willing to invest the necessary effort.

%% file: 07limitations.tex
\section{Limitations}\label{section_limitations}
Our study has threats to validity and limitations. Firstly, we acknowledge volunteer bias in our survey, as participants may be interested in replication studies. A limitation for our SLR may be one raised by the original study \cite{hao2019ReplicationSLR}, where we may have missed papers presenting a replication study because the paper did not specify replication as the study design. Another limitation is the number of years used to compare the two SLRs. The original evaluated publications over ten years, whereas our SLR used seven years, so the time periods did not align. However, as previously mentioned in \autoref{subsection_slr_results}, the original SLR did not find replication studies in the first two years, 2009-2010. Hence, the results came from eight years, bringing the time span closer.

Survey participation was voluntary and distributed through professional networks and CER venues. Hence, participants were likely more engaged and interested in research practice than the broader CER community. This could explain the relatively high proportion of participants who reported having conducted replication studies. In addition, despite the survey defining \textit{``replication''}, participants interpreted it differently, often including replication-adjacent or extension work. This, combined with the survey's reliance on participants' self-reported research histories, is likely to make the results imprecise. Participants may have underreported unsuccessful submissions, misremembered counts, or emphasised successful outcomes, particularly in relation to publication and rejection experiences.

Finally, although we applied systematic thematic analysis to the qualitative survey responses, coding necessarily involves interpretive judgment. While the themes capture recurrent patterns, alternative categorisations are possible, so these themes reflect trends in the data rather than hard boundaries or exact totals.

%% file: 08conclusion.tex
\section{Conclusions and Future Work}\label{section_conclusions}
Replication plays a crucial role in establishing consistent and reliable findings in Computing Education Research (CER). Our study offers an updated perspective on the pervasiveness and perception of replication studies within CER. By replicating a 2019 Systematic Literature Review and a survey of researchers, we found that although the CER community has historically favoured novel research over replication, there has been a modest but notable increase in the number of replication studies published between 2019 and 2025, particularly in journal venues, while conference proceedings showed a slight decrease in replication papers. This suggests that targeted initiatives, such as the CSEJ 2022 special journal issue on replication, may be effective in encouraging such work. Our survey responses indicate that, while awareness and appreciation of replication studies are improving, challenges related to recognition, publication opportunities, and perceived value remain. Overall, our results indicate progress in the acceptance and implementation of replication studies, but highlight the need for continued advocacy and support to ensure replication becomes a standard practice across all CER publication venues. Continued efforts are required to further normalise replication as a valuable research practice, especially in conferences, to help shape review practices and assist reviewers, and to support researchers undertaking this important work. Our findings underscore the importance of ongoing community dialogue and institutional support to foster a culture where replication is both valued and rewarded in CER.